\documentclass[prl,aps,superscriptaddress,fleqn,showpacs,twocolumn]{revtex4}
\usepackage{amsmath,amssymb,graphicx}
\begin{document}

\title{Lack of self-averaging in neutral evolution of proteins}

\author{Ugo~Bastolla}
\affiliation{Centro~de~Astrobiolog{\'\i}a~(INTA-CSIC),
             28850~Torrejon~de~Ardoz, Spain}

\author{Markus~Porto}
\affiliation{Max-Planck-Institut~f\"ur~Physik~komplexer~Systeme,
             N\"othnitzer~Stra{\ss}e~38, 01187~Dresden, Germany}

\author{H.~Eduardo~Roman}
\affiliation{Dipartimento~di~Fisica and INFN, Universit\`a~di~Milano,
             Via~Celoria~16, 20133~Milano, Italy}

\author{Michele~Vendruscolo}
\affiliation{Department~of~Chemistry, University~of~Cambridge,
             Lensfield~Road, Cambridge CB2~1EW, UK}

\date{March 21, 2002}

\begin{abstract}
We simulate neutral evolution of proteins imposing conservation of the
thermodynamic stability of the native state in the framework of an effective
model of folding thermodynamics. This procedure generates evolutionary
trajectories in sequence space which share two universal features for all of
the examined proteins. First, the number of neutral mutations fluctuates
broadly from one sequence to another, leading to a non-Poissonian substitution
process. Second, the number of neutral mutations displays strong correlations
along the trajectory, thus causing the breakdown of self-averaging of the
resulting evolutionary substitution process.
\end{abstract}


\maketitle

\noindent {\it Introduction.} -- The neutral theory of molecular evolution is
the simplest and most elegant theory of protein evolution. It was formulated in
the late 1960's~\cite{Kimura:1968+King/Jukes:1969} to explain the high
substitution rate of amino acids observed in proteins of many vertebrates and
the large intra-specific genetic variations between most species. The theory
assumes that most of the amino acid substitutions occurring in an evolving
population do not bring any selective advantage but are selectively neutral,
maintaining the biological activity of the protein at the original level. Such
neutral mutations are assumed to occur at random at a rate $\mu x$, where $\mu$
represents the genomic mutation rate and $x$ is the fraction of mutations which
happen to be neutral. The theory predicts that (i)~the rate of substitutions
(mutations which become fixed in the population) equals the neutral mutation
rate $\mu x$ and depends only on the protein considered, independent of the
size of the population and its ecology, and that (ii)~the number of amino acid
substitutions taking place in a time $t$ follows a Poisson distribution with
mean value $\mu x t$, thus giving an explanation to the `molecular clock'
observed in the early 1960's~\cite{Zuckerkandl/Pauling:1962}. Later studies in
the 1970's and 1980's have revealed, however, that the variance of the
substitution process is larger than its mean
value~\cite{Ohta/Kimura:1971+Langley/Fitch:1973,Gillespie:1989}, pointing to an
underlying non-Poissonian process. Since then, different alternatives have been
suggested to explain this feature. Some authors have extended the neutral
theory by including into it slightly deleterious mutations~\cite{Ohta:1976},
others have rejected the neutral theory completely and have suggested that most
mutations are fixed by positive selection~\cite{Gillespie:1991}. An interesting
proposal within the realm of neutral theory is its modification in terms of a
fluctuating neutral space model~\cite{Takahata:1987}, which can account for the
non-Poissonian statistics.

Progress in the understanding of the folding and thermodynamics of biomolecules
has opened the way to assess the thermodynamical stability of biomolecules
involved in evolution through computational methods, thus providing powerful
tools to complement the traditional population genetic approach. This
structural approach has been introduced in the study of neutral networks (i.e.,
the set of sequences connected by structure conserving mutations) of RNA
secondary structures~\cite{Schuster/etal:1994-98}, and has found fruitful
applications in the study of protein evolution~\cite{OTHERSTUDIES}. Based on
this structural approach, we have studied a model of neutral evolution denoted
{\it structurally constrained neutral model} (SCN model), in which conservation
of the thermodynamic stability of the native structure is
imposed~\cite{Bastolla/etal:2002}. The SCN model yields evolutionary
trajectories consisting of sequences connected through neutral mutations. In
this Letter, we show that the fraction of neutral mutations obtained along the
trajectories fluctuates strongly, and consequently the SCN model produces a
non-Poissonian substitution process, consistent with a fluctuating neutral
network scenario~\cite{Takahata:1987} and with the statistics of protein
evolution~\cite{Gillespie:1989}. Moreover, we find that the number of neutral
neighbors displays strong auto-correlations along the trajectories, which
deeply influence the statistics of protein evolution as we discuss below.

We have studied seven protein folds: myoglobin (Protein Database (PDB) code
{\tt 1a6g}), cytochrome~c (PDB code {\tt 451c}), lysozyme (PDB code {\tt
3lzt}), ribonuclease (PDB code {\tt 7rsa}), rubredoxin (both from a mesophilic
and a thermophilic species, PDB codes {\tt 1iro} and {\tt 1brf\_A}), ubiquitin
(PDB code {\tt 1u9a\_A}), and the TIM barrel (PDB code {\tt
7tim\_A})~\cite{Bastolla/etal:2002}. Although these proteins cover a broad
spectrum of different biological activities, according to the SCN model their
neutral evolution occurs in a rather similar way, suggesting that our model
captures `universal' features of protein evolution. In what follows, we
concentrate on myoglobin to illustrate this common scenario.

\noindent {\it The SCN
model\/}~\cite{Bastolla/etal:2002,Bastolla/etal:1999+2000}. -- The simulated
trajectories are started from a given ``wild-type'' sequence ${\bf A^*}$ of $N$
amino acids (residues) that folds onto the native structure ${\bf C^*}$, where
both ${\bf A^*}$ and ${\bf C^*}$ are taken from the PDB. The target structure
${\bf C^*}$ is kept fixed throughout the simulation. Following the neutral
theory, a given mutated sequence ${\bf A'}$ is considered to be either neutral,
if it folds onto ${\bf C^*}$ preserving thermodynamic stability, or lethal
otherwise. Neither advantageous nor slightly deleterious mutations are
considered. At each iteration, we generate all possible sequences $\{ {\bf A'}
\}$ obtained through point mutations of the current sequence ${\bf
A}$~\cite{Note}, and determine the fraction belonging to the neutral network.
One of these neutral sequences is chosen at random and becomes the new current
sequence. The whole process is iterated, resulting in an evolutionary
trajectory $\{ {\bf A}_0 = {\bf A^*}, {\bf A}_1, {\bf A}_2, \ldots \}$.

To assess the conservation of the thermodynamic stability of a test sequence in
the target structure, we rely on two well-established empirical parameters, the
normalized energy gap $\alpha$~\cite{Bastolla/etal:1999+2000}, measuring the
minimal value of the energy gap between an alternative conformation and the
target one divided by their structural distance, and the
$Z$-score~\cite{Bowie/etal:1991+Goldstein/etal:1992}, measuring the difference
between the native energy and the average energy of alternative compact
conformations in units of the standard deviation of the energy. A positive and
large value of the $\alpha$-parameter ensures both that the target conformation
has lowest energy and that the energy landscape is well correlated, the latter
in the sense that conformations very different from the native have much higher
energy. Moreover, the native energy allows a rough estimate of the folding free
energy. Alternative structures are generated by aligning in all possible ways
the test sequence with all non-redundant protein structures in the PDB.
Conformational energies are computed through the effective energy function
described in Ref.~\cite{Bastolla/etal:2000+Bastolla/etal:2001}. The resulting
estimate of the thermodynamic stability provides a realistic genotype to
phenotype mapping, although the energy function used is approximate and
alternative structures are non-exhaustively sampled.

\begin{figure}[t]
\centerline{\includegraphics[scale=0.53]{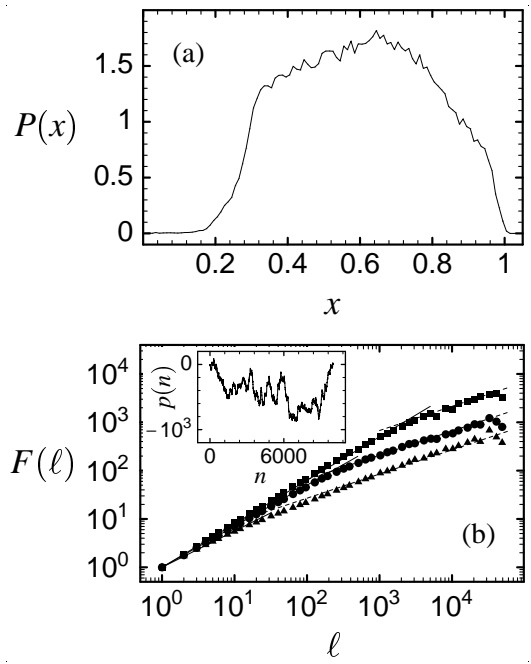}}
\caption{(a)~Probability distribution $P(x)$ of the fraction $x$ of neutral
neighbors for myoglobin. (b)~Analysis of the fluctuations of $x$ along an
evolutionary trajectory of myoglobin, displayed as $F(\ell)$ vs $\ell$. We
present the analysis of the whole protein (triangles) and the analysis for
residues $25$ (circles) and $115$ (squares). The full lines have the slopes
$0.82$ (for the triangles, up to $\ell \approx 20$), $0.85$ (for the circles,
up to $\ell \approx 10^2$), and $0.9$ (for the squares, up to $\ell \approx
10^3$) indicating strong correlations, whereas the dashed lines have the slope
$1/2$, indicating the absence of correlations. The inset in (b) shows part of
the profile $p(n)$ vs $n$ for the total fraction of neutral neighbors (see
text).}
\label{figure:Px+Fl}
\end{figure}

\noindent {\it Neutral connectivity.} -- In the SCN model, the quantity of
interest is the fraction of neutral neighbors $x_i = X_i/X_{\rm max} \in (0,1]$
associated to each sequence ${\bf A}_i$, as this quantity determines the
evolutionary substitution process (see below). Here, $X_i$ is the number of
neutral neighbors of sequence ${\bf A}_i$, and $X_{\rm max}$ is its maximal
possible value~\cite{Note}. Thus, an evolutionary trajectory is an (ideally
infinite) series ${\bf x} = \{ x_0, x_1, x_2, \ldots \}$. For each position
$k$, we define analogously evolutionary trajectories ${\bf x}^{(k)}$, where
$x_i^{(k)}$ counts the fraction of the $18$ possible changes at position $k$
which are neutral. The probability distribution $P(x)$, where $P(x) \, {\rm
d}x$ is the probability to find a fraction of neutral neighbors between $x$ and
$x+{\rm d}x$ in the trajectory ${\bf x}$, obtained for myoglobin is shown in
Fig.~\ref{figure:Px+Fl}(a). The fraction of neutral neighbors is broadly
distributed, reminiscent of a fluctuating neutral network~\cite{Takahata:1987}.

To get a deeper insight into the mechanism of neutral evolution, it is helpful
to investigate the auto-correlation of the trajectory ${\bf x}$. For this
purpose, we study the `profile' $p(n)$ of walks of $n$ steps, defined as $p(n)
= \sum_{i=1}^n \left[ x_i-\overline{x} \right]$, where $\overline{x} = \int_0^1
x \, P(x) \, {\rm d}x$ is the average over all $x_i$ in the trajectory (cf.\
inset in Fig.~\ref{figure:Px+Fl}(b)). The `roughness' of the profile tells us
about the presence or absence of correlations along the trajectory ${\bf x}$.
They can be determined quantitatively by calculating the fluctuations of $p(n)$
within a window of width $\ell$ as $F(\ell) = \big< [p(n) - p(n+\ell)]^2
\big>^{1/2}$ (see Ref.~\cite{Peng/etal:1994+Koscielny-Bunde/etal:1998}). If
${\bf x}$ is a series of independent variables, $F(\ell)$ scales as
$\ell^{1/2}$. Instead, if the series ${\bf x}$ is correlated (anti-correlated)
over a length $\ell_0$, the scaling is modified into $F(\ell)\sim \ell^{\nu}$
with $\nu > 1/2$ ($\nu < 1/2$) on the same length scale. The information
provided by $F(\ell)$ is equivalent to the one given by the auto-correlation
function, but the former has the advantage that different regimes and
intermediate crossovers are more easily detected.

The plot of the fluctuations of an evolutionary trajectory ${\bf x}$ obtained
for myoglobin is shown in Fig.~\ref{figure:Px+Fl}(b), together with the
analogous analysis of the fraction of neutral neighbors for residues $25$ and
$115$, respectively the residues with the largest and the smallest mean
fraction of neutral neighbors. The fraction of neutral neighbors for the whole
protein is strongly auto-correlated, with an apparent exponent $\nu \approx
0.82$ for about $20$ evolutionary steps, after which the random walk exponent
$\nu = 1/2$ is attained. Single residues have a somewhat larger apparent
exponent between $\nu \approx 0.85$ (residue $25$) and $\nu \approx 0.9$
(residue $115$) for about $10^2$ to $10^3$ evolutionary steps, which is roughly
the product of number of steps it takes the correlations to vanish for the
whole protein times the protein length. These auto-correlations can be
qualitatively explained by the fact that more stable sequences have a larger
number of neutral neighbors (they are more tolerant to mutations), and
stability itself is auto-correlated along an evolutionary trajectory as long as
the number of mutations is small with respect to sequence length. These
correlations, although being short range, have an important influence on the
evolutionary substitution process, as shown below.

\noindent {\it Modeling the substitution process.} -- In order to construct the
substitution process, we have to obtain the number of mutations accepted within
a time interval $t$. The substitution process consists of three steps:
(i)~generation of an evolutionary trajectory using the SCN model;
(ii)~determination of the number $k$ of mutation events taking place within the
time $t$, which is assumed to be a Poissonian variable of average $\mu t$, so
that the probability of $k$ mutations within time $t$ is $P_t(k) = \exp(-\mu t)
\, (\mu t)^k/k!$; (iii)~determination of the number $n$ of accepted mutation
events out of $k$, where the corresponding conditional probability $P_{\rm
acc}(n \mid k) = \left[ \prod_{i=1}^{n} x_i \right] \sum_{\{ m_j \}}
\prod_{j=1}^{n+1} (1-x_j)^{m_j}$. Here, the $\{ m_j \}$ are all integer numbers
between zero and $k-n$ satisfying $\sum_{j=1}^{n+1} m_j = k-n$. In other words,
the probability that a mutation is accepted is $x_0 = x({\bf A}_0)$ as long as
the protein sequence is ${\bf A}_0$, $x_1 = x({\bf A}_1)$ as long as the
sequence is ${\bf A}_1$, and so on. If all $x_i = x$ are equal, $P_{\rm acc}(n
\mid k)$ reduces to a binomial distribution $P_{\rm acc}(n \mid k) =
\binom{k}{n} \, x^n \, (1-x)^{k-n}$. The probability $\Pi_t(n)$ that $n$
mutations are accepted within time $t$ is the weighted sum over $k$ of the
acceptance probability, $\Pi_t(n) = \sum_{k \ge n} P_t(k) \, P_{\rm acc}(n \mid
k)$, which in the case of equal $x_i$ reduces to a simple Poisson distribution
$\Pi_t(n) = \exp(-\mu t x) \, (\mu t x)^n/n!$ with average value $\mu x t$ and
substitution rate $\mu x$, as in the original neutral model.

\begin{figure}[t]
\centerline{\includegraphics[scale=0.53]{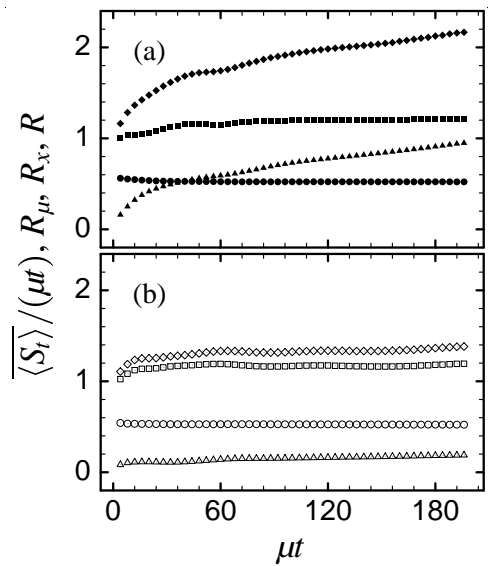}}
\caption{Statistics of the substitution process for myoglobin, shown are the
average number of substitutions $\overline{\big< S_t \big>}$ divided by $\mu t$
(circles), the normalized mutation variance $R_{\mu}$ (squares), the normalized
trajectory variance $R_x$ (triangles), and the normalized total variance $R =
R_{\mu}+R_x$ (diamonds). We present in (a) the statistics for the trajectories
obtained through the SCN model (full symbols), and in (b) the same for the
corresponding annealed trajectories (open symbols).}
\label{figure:substitution}
\end{figure}

\noindent {\it Statistics of the substitution process.} -- In our analysis, two
kinds of averages must be distinguished. We indicate by angular brackets $\big<
\cdot \big>$ the average over the mutation and acceptance process for a given
realization of the evolutionary trajectory, and by an overline
$\overline{\vphantom{S}\cdot\vphantom{S}}$ the average over evolutionary
trajectories. We determine the mean number of substitutions $\overline{\big<
S_t \big>}$ within a time $t$, and show this quantity in
Fig.~\ref{figure:substitution} together with the normalized mutation variance
$R_{\mu} =(\overline{\big< S_t^2 \big>}-\overline{\big< S_t \big>^2})/
\overline{\big< S_t\big>}$, the normalized trajectory variance $R_x =
(\overline{\big< S_t \big>^2}-\overline{\big< S_t \big>}^2)/\overline{\big< S_t
\big>}$, and the normalized total variance $R = R_{\mu}+R_x$. The latter
quantity is also known as dispersion index. Notice that if all $x_i = x$ are
the same, one gets $R_{\mu} = 1$ as for all Poissonian processes, and the
normalized trajectory variance $R_x = 0$, since $x$ is constant in all sequence
space.

The results of the substitution process based on the evolutionary trajectories
are shown in Fig.~\ref{figure:substitution}(a). We also show in
Fig.~\ref{figure:substitution}(b) results based on {\it annealed} trajectories,
obtained by extracting at random the values of $x_i$ at each substitution event
according to the observed distribution $P(x)$. In this case, the different
$x_i$ along the trajectories are independent variables. Note that the annealed
trajectories `interpolate' between the Poissonian case (all $x_i$ are equal)
and the correlated trajectories obtained through the SCN model: (i)~The
comparison between the annealed trajectories and the simple Poissonian case
allows us to identify the effect of the broad distribution of the fraction of
neutral neighbors, whereas (ii)~the comparison between the actual and the
annealed trajectories allows us to identify the effect of correlations.

In the annealed case, the time spans $\tau$ between subsequent substitutions
are independent variables distributed with the density $\overline{D(\tau)} =
\int_0^1 P(x) \, (\mu x)^{-1} \exp(-\mu x \tau) \, {\rm d}x$, whose average
value is $\overline{\tau} = \int_0^1 P(x) \, (\mu x)^{-1} \, {\rm d}x$. Thus,
the average substitution rate $\overline{\big< S_t \big>}/t$ is not constant in
time as in the Poissonian case. Initially, $S_t$ is a Poissonian variable with
average rate $\mu \overline{x}$. At large time, however, the rate converges to
the smaller value $\overline{\big< S_t \big>}/t \approx 1/\overline{\tau}$,
since the process spends more and more time in sequences with small $x$. Hence,
the standard deviation of $\tau$ is slightly larger than its average value, and
the normalized mutation variance $R_{\mu}$ is larger than the Poissonian value
$R_{\mu} = 1$, although the actual difference is small (see
Fig.~\ref{figure:substitution}(b)). The normalized trajectory variance $R_x$ is
very small, of the order of the ratio between variance and average value of the
$x_i$.

Using the actual evolutionary trajectories (see
Fig.~\ref{figure:substitution}(a)), we note that the presence of correlations
has only a weak effect both on the average number of substitutions
$\overline{\big< S_t \big>}$ and on the normalized mutation variance $R_{\mu}$.
However, the normalized trajectory variance increases considerably in response
to the correlations, as $R_x \approx 1$ for $\mu t = 200$. It even grows with
time, although more and more sequences are used to compute the mutational
averages and one could naively expect that such averages approach typical
values. Hence, the large fluctuations between different trajectories caused by
the strong auto-correlations result in the breakdown of self-averaging in the
substitution process, in the sense that even averaging over an arbitrary long
trajectory does not give values representative of typical trajectories. (We
note, however, that the variable $\big< S_t \big>/t$ is still self-averaging as
its variance vanishes in the long $t$ limit, but the normalized variable $\big<
S_t \big>/\sqrt{t}$ has a variance which increases with time.) The normalized
total variance $R$ becomes larger than $2$ already for $\mu t \approx 100$, in
agreement with empirical estimates, varying in the range $1 < R \lesssim 5$ for
most of the proteins
studied~\cite{Ohta/Kimura:1971+Langley/Fitch:1973,Gillespie:1989}. Hence, these
large dispersion indices may be to a large extent due to the correlations
present in the evolutionary process.

\noindent {\it Conclusions.} -- We have shown that the evolutionary
trajectories in sequence space generated by the SCN model are characterized
both by a broad distribution of neutral connectivities and by strong
correlations, which deeply influence the statistics of neutral evolution and
the substitution process. For example, fluctuations of the evolutionary rate
from one branch of the evolutionary tree to the other can obscure lineage
effects, i.e.\ variations of the substitution rate among different taxonomic
groups. One such effect is the generation time effect: Since the natural time
unit for measuring substitution events is the generation time (at which
reproduction takes place), the longer the latter the slower the substitution
rate is expected to be. This has been verified by comparing for instance
substitution rates in rodents and
primates~\cite{Britten:1986+Li/etal:1987+Grauer/Li:2000}. However, the effect
is significantly larger for synonymous substitutions (DNA changes which still
code for the same amino acid) than for non-synonymous ones. Non-synonymous
substitutions are superimposed with the large and correlated fluctuations in
the substitution rate that we just described, while synonymous ones are not.
Thus the statistics of neutral substitutions could explain this quantitative
difference. Additionally, a better understanding of the mechanisms of neutral
evolution will help to single out the more interesting cases of positive
selection as, for instance, changes in the protein function and responses to
variations of the environment. The best current bioinformatics method to
identify such cases of positive selection~\cite{McDonald/Kreitman:1991},
recently used to study the evolution of {\it Drosophila}
genes~\cite{Smith/Eyre-Walker:2002+Fay/etal:2002}, assumes a neutral
substitution process with constant fraction of neutral neighbors $x$. The broad
distribution and the correlations that we observe can mimick the presence of
positive selection, and they should be taken into account to improve the
performance of such methods.


\begin{thebibliography}{99}
\bibitem{Note}
 We impose conservation of the cysteine residues and do not allow any residue
 to mutate to a cysteine, since a mutation changing the number of cysteine
 residues by one would leave the protein with a very reactive unpaired
 cysteine. Accordingly, the total number of neighbors is $X_{\rm max} = 18
 (N-N_{\rm cys})$, where $N_{\rm cys}$ is the number of cysteine residues in
 the starting sequence. For instance for myoglobin, one has $N = 151$, $N_{\rm
 cys} = 0$, and hence $X_{\rm max} = 2718$.
\end{thebibliography}
\end{document}